\documentclass[11pt,english]{article}
\usepackage{amsfonts, amsmath, amssymb, verbatim, setspace, pdfsync, graphicx}
\usepackage{comment}
\usepackage{color}
\usepackage{pdfsync}
\definecolor{darkgreen}{rgb}{0,0.5,0}
\definecolor{darkblue}{rgb}{0,0,0.6}
\definecolor{purple}{rgb}{0.4,0.15,0.21}
\definecolor{black}{rgb}{.2,.2,.2}
\usepackage[colorlinks=true,citecolor=darkblue,linkcolor=black,urlcolor=darkblue]{hyperref}

\DeclareMathOperator{\p}{\partial}
\DeclareMathOperator{\h}{\theta}

\DeclareMathOperator{\Tr}{Tr}

\newcommand{\be}{\begin{equation}}
\newcommand{\ee}{\end{equation}}

\newcommand{\f}{\frac}

\begin{document}
\unitlength = 1mm
\

\begin{center}


{ \LARGE {\textsc{\begin{center}Black hole microstates in AdS \end{center}}}}

\vspace{0.8cm}
Edgar Shaghoulian

\vspace{.5cm}

{\it Department of Physics} \\
{\it University of California}\\
{\it Santa Barbara, CA 93106 USA}

\vspace{1.0cm}

\end{center}

\begin{abstract}
\noindent We extend a recently derived higher-dimensional Cardy formula to include angular momenta, which we use to obtain the Bekensten-Hawking entropy of AdS black branes, compactified rotating branes, and large Schwarzschild/Kerr black holes. This is the natural generalization of Strominger's microscopic derivation of the BTZ black hole entropy to higher dimensions. We propose an extension to include $U(1)$ charge, which agrees with the Bekenstein-Hawking entropy of large Reissner-Nordstrom/Kerr-Newman black holes at high temperature. 
We extend the results to arbitrary hyperscaling violation exponent (this captures the case of black D$p$-branes as a subclass) 
and reproduce logarithmic corrections. 

\end{abstract}

\pagebreak
\setcounter{page}{1}
\pagestyle{plain}

\setcounter{tocdepth}{1}

\tableofcontents

\section{Introduction}\label{intro}
In a recent paper \cite{Shaghoulian:2015kta}, a formula was presented which relates the entropy of a conformal field theory (CFT) at high temperature to the vacuum energy on $S^1 \times \mathbb{R}^d$: 
\be\label{thermo}
S= (d+1)T^d V_d\, \varepsilon_{\textrm{vac}}\,.
\ee
The number $\varepsilon_{\textrm{vac}}$ is defined as $E_{\textrm{vac}} = -\varepsilon_{\textrm{vac}} V_d/L^{d+1}$ where $L$ is the circumference of the $S^1$. The periodicity conditions along the $S^1$ match the thermal periodicity conditions. In particular, fermions will be antiperiodic along the spatial $S^1$. 
 The entropy calculated admits a microscopic interpretation as a degeneracy of states at asymptotically large energy:
\be\label{micro}
\log \rho(E) = \f{d+1}{d^{\f{d}{d+1}}}\; (\varepsilon_{\textrm{vac}} V_d)^{\f{1}{d+1}} E^{\f{d}{d+1}}\,.
\ee
For $d=1$, these formulas reduce to the famous Cardy formula of two-dimensional CFTs \cite{Cardy:1986ie}. Picking the spatial background to be $S^1\times \mathbb{R}^d$ allows the cleanest derivation of these results in terms of the approximate modular invariance on this background:
\be\label{modular}
\log Z(\beta) = \left(\f{L}{\beta}\right)^{d-1}\log Z\left(\f{L^2}{\beta}\right)\,.
\ee

There is also a generalized entropy formula in \cite{Shaghoulian:2015kta} which incorporates a hyperscaling violation exponent $\h$:
\be\label{thermoeff}
S =- (d_{\textrm{eff}}+1)  \,L^{d_{\textrm{eff}}+1}\,T^{d_{\textrm{eff}}} \,E_{\textrm{vac}}\,,
\ee
where $d_{\textrm{eff}}=d-\h$. This can also be understood as a microcanonical degeneracy of states:
\be\label{microeff}
\log \rho(E) = \f{d_{\textrm{eff}}+1}{d_{\textrm{eff}}\,^{\f{d_{\textrm{eff}}}{d_{\textrm{eff}}+1}}}\; (-E_{\textrm{vac}})^{\f{1}{d_{\textrm{eff}}+1}} E^{\f{d_{\textrm{eff}}}{d_{\textrm{eff}}+1}}L\,.
\ee
 
We will use these formulas to account for the Bekenstein-Hawking entropy of the following black brane solutions:
\begin{align}\label{blackhole}
ds^2_{d+2}=\ell^2r^{-2\h/d}\left(-r^2\left(1-(r_h/r)^{d+1-\h}\right)dt^2+\f{dr^2}{r^2\left(1-(r_h/r)^{d+1-\h}\right)}+r^2dx_i^2\right),
\end{align}\vspace{-7mm}
\begin{align}\label{bh}
S=\f{A}{4G} = \f{V_d \,\ell^d r_h^{d-\h}}{4G }\,, \qquad \quad \beta = \f{4\pi}{ r_h(d+1-\h)}\,.
\end{align}
The parameter $V_d$ represents the (infinite) area of the black brane horizon. The case $d=1$ is discussed in \cite{Shaghoulian:2015dwa, Bravo-Gaete:2015wua}. The vacuum energy necessary to apply the formula is postulated to be given by the gravitational soliton obtained by a double Wick rotation of the black brane. This soliton gives the \emph{strongly coupled} vacuum energy, which is necessary to produce the strongly coupled entropy. The energy of the gravitational soliton is straightforwardly computed by the Balasubramanian-Kraus procedure of holographically renormalizing the Brown-York stress-energy tensor \cite{Balasubramanian:1999re}. 

We will begin with $\h=0$ and $d>1$, which corresponds to AdS black branes in arbitrary dimension. This contains as a subcase black D$3$-branes. The gravitational soliton used will be none other than the AdS soliton. An argument that this configuration locally minimizes the energy was provided in \cite{Horowitz:1998ha}, while evidence that the soliton provides a global ground state for the given boundary conditions was provided in \cite{Galloway:2001uv}.  After treating this case, we will connect our  formalism to that of Euclidean gravity in section \ref{euclidean} to illustrate why it is guaranteed to work. 

In section \ref{angularsec}, we will derive field-theoretic extensions to the higher-dimensional Cardy formulas that include angular momentum on a torus and on a sphere. We will show that the resulting formulas reproduce the entropy of compactified boosted black branes and large/hot AdS-Kerr black holes, respectively. We will also present a new derivation of the Cardy formula at finite angular momentum for two-dimensional CFTs and connect it to the Hartman-Keller-Stoica conjecture for the universality of the free energy in the range $E_LE_R>(c/24)^2$ \cite{Hartman:2014oaa}.

Another natural extension is to include a global $U(1)$ charge. In this case, we propose a  modular behavior of the grand canonical partition which mimics what happens in two dimensions. With this modular behavior, we are able to derive a formula for the entropy as an expansion in small charge for three and four spacetime dimensions, which agrees with the entropy of large AdS-Reissner-Nordstrom black holes in four and five bulk spacetime dimensions. The case of both $U(1)$ charge and angular momentum (either on the torus or the sphere) immediately follows from the techniques of section \ref{angularsec}. These field theory formulas are then shown to agree with compactified boosted AdS-Reissner-Nordstrom black branes and large/hot AdS-Kerr-Newman black holes. Finally, we will outline the matching in the case $\h \neq 0$ (which contains as a subclass black D$p$-branes for $p \neq 3$) by appealing to the methods of section \ref{euclidean}.

\section{AdS black branes and large AdS-Schwarzschild black holes}
Although we can immediately perform the analysis in general dimension, we will for clarity consider a few individual cases before giving the general answer. There are two ways one can interpret the application of the formulas of the previous section. The first is to consider an asymptotically locally AdS spacetime where one of the spatial coordinates is compactified and the fields have thermal periodicity conditions along this spatial coordinate. This is the more canonical setup since the ground state soliton and the black brane are in the same asymptotic class of data. In other words, both configurations are describing the field theory on the same background with the same periodicity conditions. There is also a Hawking-Page transition between the thermal soliton and the black brane at $\beta = L$, where $L$ is the periodicity of the spatial circle. This is the self-dual point of the modular transformation \eqref{modular}, just as in $1+1$ dimensions. The second way to interpret the application of the formula is to consider asymptotically AdS spacetime with all coordinates non-compact. The entropy of the black brane in this setup is then related to the ground state soliton, which is in a different class of asymptotic data. From the field theory point of view, the entropy of the field theory on the plane $\mathbb{R}^d$ is related to the ground state energy on $S^1 \times \mathbb{R}^{d-1}$ where the fields have thermal periodicity conditions along the $S^1$. Both interpretations are consistent with the calculations below and the difference enters only into the volume factor $V_d$. 

\subsection{AdS$_{d+2}$}
We will start with three dimensions and move up to the case of general dimension. In all cases the vacuum energy at strong coupling is assumed to be given by the energy of the AdS soliton:
\be
ds^2=\ell^2\left(-r^2dt^2+r^2dx_i^2+r^2\left(1-(\tilde{r}_h/r)^{d+1}\right)d\phi^2+\f{dr^2}{r^2\left(1-(\tilde{r}_h/r)^{d+1}\right)}\right).
\ee
This geometry is produced by a double analytic continuation from the AdS black brane. Assuming that the $\phi$ cycle is of length $L$ fixes $\tilde{r}_h = (d+1)L/4\pi$. The energy of this spacetime can be calculated by the usual Balasubramanian-Kraus procedure of regularizing the Brown-York stress-energy tensor, and gives 
\be
\varepsilon_{\textrm{vac}}=\f{\ell^d}{16\pi G}\left(\f{4\pi}{d+1}\right)^{d+1}\,.
\ee
\subsubsection{AdS$_3$}
This is simply the case of a BTZ black hole in the bulk, and our formula reduces to the usual Cardy formula of $(1+1)$-dimensional CFT. The gravitational soliton obtained by swapping thermal and angular cycles is global AdS$_3$. The entropy is correctly reproduced as shown in \cite{Strominger:1997eq}.

\subsubsection{AdS$_4$}
The vacuum energy and the black brane energy are given by
\be
\varepsilon_{\textrm{vac}} = \f{4\pi^2 \ell^2}{27 G }\,,\qquad E = \f{\ell^2 V_2r_h^3}{8 \pi G }\,.
\ee
Notice that $\varepsilon_{\textrm{vac}}$ is independent of $L$, the periodicity of $\phi$. Thus, as a practical matter, the periodicity is irrelevant as long as there are no conical defects. A simple choice is to pick periodicity $\beta$, in which case the soliton is most simply related to the black brane. The more canonical setup is to begin with a black brane on a spatial background with $\phi$ of a given periodicity, and ensure that $\phi$ has the same periodicity for the soliton. This keeps the analysis within a fixed set of asymptotic data.  

Plugging the above energies into the microcanonical degeneracy \eqref{micro} gives
\be
S=\f{V_2 \ell^2 r_h^2}{4G}\,.
\ee
This is precisely the Bekenstein-Hawking entropy \eqref{bh} of the black brane. We could have equivalently plugged the temperature of the black brane and $\varepsilon_{\textrm{vac}}$ into \eqref{thermo} to achieve the same result.

\subsubsection{AdS$_5$ and D$3$-branes}\label{genmodular}
The same renormalization procedure gives 
\be
\varepsilon_{\textrm{vac}} =\frac{\pi ^3 \ell^3 }{16 G},\qquad E =\f{3\ell^3 V_3 r_h^4}{16\pi G }\,.
\ee
Plugging into \eqref{micro} gives the correct Bekenstein-Hawking entropy. In this example, we can translate into gauge theory parameters corresponding to the example of a stack of D$3$-branes. In that case, we have 
\be
\varepsilon_{\textrm{vac}} = \f{\pi^2N^2}{8}\,,
\ee
as first showed in \cite{Horowitz:1998ha}. Formula \eqref{thermo} relates this to the entropy of the gauge theory by 
\be
S = 4 V_3 T^3 \varepsilon_{\textrm{vac}}\,.
\ee
Since this equation is true at strong and weak coupling, we see explicitly that the factor of $3/4$ between the strongly coupled and weakly coupled thermal entropies must manifest itself in the Casimir energies, as first explained from the gravity point of view by \cite{Horowitz:1998ha}. 

\subsubsection{AdS$_{d+2}$}
The energy of the AdS soliton and black brane is given in general dimension as 
\be\label{generald}
\varepsilon_{\textrm{vac}} = \f{\ell^d}{16 \pi G}\left(\f{4\pi}{d+1}\right)^{d+1}, \qquad E =\f{d\ell^d V_d  r_h^{d+1}}{16\pi G}\,.
\ee
Using these expressions gives the correct entropy.

\subsection{AdS-Schwarzschild black hole and local operators}
At large temperature, the entropy of the AdS-Schwarzschild black hole becomes that of the AdS black brane. In this limit, as discussed in \cite{Shaghoulian:2015kta}, the formula \eqref{micro} with $V_d =\textrm{Vol}(S^d)= 2\pi^{(d+1)/2}/\Gamma[(d+1)/2]$ can be understood as a degeneracy of local operators. This then reproduces the entropy of the AdS-Schwarzschild black hole in terms of a count of the local operators of the dual CFT. 

\subsection{Additional compactifications}
A natural question, first explored in \cite{Myers:1999psa}, regards representing the supergravity backgrounds dual to the ground state of the CFT under additional spatial compactifications. For this discussion we will assume all spatial directions are compactified onto a torus. The derivation of our entropy formula in \cite{Shaghoulian:2015kta} sheds some light on this issue. Recall that in that derivation one begins with the partition function of the field theory at finite temperature. If there is a spatial direction much larger than the rest, then we choose it as the thermal direction and an equivalent representation of the partition function can be presented. This is approximately the zero temperature partition function on a new torus where the original large spatial cycle was swapped for a cycle of length $\beta$. In this representation, it is clear that the vacuum state makes the dominant contribution to the partition function. When $\beta$ is very small the bulk representation is the black brane. The ground state energy is for the theory on a manifold which is effectively $S^1 \times \mathbb{R}^{d-1}$, since the $\beta$ cycle is much smaller than the rest. The AdS soliton gives the correct energy for the field theory on such a background (correct in the sense that it is connected to the entropy of the black brane by \eqref{thermo}). 

Now let us assume that $\beta$ is smaller but comparable to the cycles of the spatial torus and there exists a spatial cycle much larger than the rest. If the black brane remains the dominant configuration to this ensemble, then by the same argument of swapping cycles one can be sure that the AdS soliton with additional compactifications represents the ground state. The formula for the entropy in terms of the ground state energy will continue to be satisfied simply because it was satisfied for the situation of asymptotically small $\beta$. A similar perspective was presented in \cite{Myers:1999psa}.

\section{Comparison to Euclidean gravity}\label{euclidean}
Obtaining the entropy from a Euclidean gravity calculation and from the method of the previous section share a common feature: both methods localize the partition function on a single contribution. We will now show that they both localize to essentially the same piece, so our derivation can be understood as providing a microscopic understanding of the methods of Euclidean gravity in the regime where both methods are applicable. 

Recall that the method of Euclidean gravity begins with a path integral which localizes on a classical saddle:
\be
Z(\beta) = \int [\mathcal{D}g] \;e^{-I_E} \approx e^{-I_E^{\textrm{black hole}}}\,.
\ee
We assume that $\beta\ll L$ for some spatial cycle $\phi$ of size $L$. Now consider the soliton obtained by the double Wick rotation $\{t,\phi\}\rightarrow \{i\phi, it\}$. For the Euclidean geometry to be free of deficits, we need $\phi\sim \phi+\beta$. This is the same Euclidean geometry as the black hole, so we have
\be
Z(\beta) \approx e^{-I_E^{\textrm{soliton}}} = e^{-L F}\,.
\ee
Since the soliton has zero classical entropy, its free energy equals its ordinary energy and we have 
\be
Z(\beta)\approx e^{-L E_{\textrm{soliton}}}\,.
\ee
This is precisely the expression we get by swapping cycles in our field theory derivation:
\be
Z(\beta) = \sum e^{-\beta E_{L\times \mathbb{T}^{d-1}}} = \sum e^{-L E_{\beta\times \mathbb{T}^{d-1}}} \approx e^{-L E_{\textrm{vac}, \beta\times \mathbb{T}^{d-1}}}\,.
\ee
The subscript on the energy denotes the spatial manifold of the field theory. We now see that identifying $E_{\textrm{vac}, \beta\times \mathbb{T}^{d-1}}$ with the energy of the bulk soliton is guaranteed to work since the Euclidean gravity method gives the correct entropy.

When we actually apply our formula in the context of holography, we may want to ensure that the soliton is in the same class of asymptotic data as the black brane. In that case, the field theory argument of \cite{Shaghoulian:2015kta} requires performing a scale transormation to restore the size of the spatial circle. In the argument above, that would mean performing a scale transformation such that $\phi$ rescales by $L/\beta$. For simplicity, consider the case $\h=0$. In that case, the bulk scale transformation simply changes $r_h \rightarrow (L/\beta)r_h$. This uniquely fixes $r_h$ given a fixed periodicity of $\phi$ (i.e. $r_h$ is independent of the temperature of the black brane one began with). This rescaling is what gives the ordinary AdS soliton. It has a different Euclidean action than (and participates in a Hawking-Page transition with) the black brane.

These arguments explain the successful use of solitons in accounting for black hole entropy in the literature.

\section{Angular momentum}\label{angularsec}
We can also consider the degeneracy of states at finite energy and finite angular momentum. We will begin by deriving the field theory formula with angular momentum on a torus background in two different ways in sections \ref{take1} and \ref{take2}. The appropriate gravity background to compare to is a compactified black brane with rotation along the compact direction.  These backgrounds can be constructed by first boosting the non-compact, non-rotating branes and then compactifying them. (In three spacetime dimensions, this boosting procedure produces the rotating BTZ black hole from the non-rotating case \cite{Clement:1995zt, Martinez:1999qi}.) In section \ref{boostbrane} we will find that the Bekenstein-Hawking entropy of these boosted branes is precisely reproduced by the logarithm of the microcanonical degeneracy of states at finite energy and angular momenta of the dual field theory. 

The case of a conformal field theory on $S^1 \times S^d$ with $\left \lfloor{(d+1)/2}\right \rfloor$ independent rotation parameters along the $S^d$  is treated in section \ref{sphereft}. Working in the fluid-dynamical regime, the partition function of the theory with angular momentum is simply related to the partition function of the theory without angular momentum. Assuming as usual that the theory on the sphere at high temperature can be replaced by the theory on the plane (i.e. the curvature of the $S^d$ becomes irrelevant for small $S^1$), we can then relate the case with angular momenta on the sphere to the case with vanishing momentum on a plane or torus. The appropriate gravity background to compare to is a large Kerr-AdS black hole. In section \ref{spheregravity} we will find that the Bekenstein-Hawking entropy of the Kerr-AdS black hole in any number of dimensions with any number of independent rotation parameters is precisely reproduced by the logarithm of the microcanonical degeneracy of states at finite energy and angular momenta of the dual field theory. 

\subsection{Angular momentum on the torus: old-school derivation}\label{take1}
We need to generalize the field theory derivation of the degeneracy of states with energy $E$ in \cite{Shaghoulian:2015kta} to include angular momentum. The partition function for this ensemble is given as $Z(\beta, \h) = \Tr e^{-\beta E+i\h J}$ for inverse temperature $\beta$ and angular potential $\h$. 
We will restrict the angular momentum to be in one direction. The general case can be accommodated by the methods below.

We proceed by using the modular properties of \cite{Shaghoulian:2015kta} on $\mathbb{T}^2 \times \mathbb{T}^{d-1}$, where the $\mathbb{T}^{d-1}$ is taken to be large. The modular parameter of the torus $\mathbb{T}^2$ is taken to be $2\pi\tau = 2\pi r e^{i\phi}=\beta+i\theta$ and $2\pi\bar{\tau} = 2\pi\tau^* = 2\pi r e^{-i\phi} = \beta-i\h$. Notice this means we are not taking $\tau$ and $\bar{\tau}$ to be independent coordinates. We can write the partition function as 

\begin{align}
Z(\tau,\bar{\tau}) =  \Tr &\left(e^{2\pi i \tau E_R}e^{-2\pi i \bar{\tau}E_L}\right)=Z(-1/\tau,-1/\bar{\tau})^{r^{1-d}} ,\\
&E_R+E_L = E, \qquad E_R-E_L = J\,.
\end{align}
The microcanonical degeneracy of states is obtained by an inverse Laplace transform
\be
\rho(E,J) = \int dr\, d\phi \,Z(r,\phi)\; \exp\left[-2\pi i r e^{i\phi} E_R+2\pi i re^{-i\phi}E_L\right].
\ee
Defining 
\be
\widetilde{Z}(r,\phi) = \Tr\left[ \exp\left(2\pi i r e^{i\phi}(E_R-E_{\textrm{vac}}/2)-2\pi i r e^{-i\phi}(E_L-E_{\textrm{vac}}/2)\right)\right] 
\ee
and using the modular transformation of $Z$ gives 
\be
\rho(E,J)=\int dr\, d\phi \,\widetilde{Z}\left(-\f{1}{r},-\phi\right)^{r^{1-d}} \exp\left[\f{-\pi i E_{\textrm{vac}}}{r^de^{i\phi}}+\f{\pi i E_{\textrm{vac}}}{r^de^{-i\phi}}-2\pi i r e^{i\phi}E_R+2\pi i re^{-i\phi}E_L\right].
\ee
This can be evaluated by saddle-point approximations for $r$ and $\phi$, where we treat $\widetilde{Z}$ as a slowly varying prefactor. In the limit $E_{R,L}\gg |E_{\textrm{vac}}|$ the saddle-point values are
\begin{align}
r_s=\left(\f{-E_{\textrm{vac}}(d-1)\left(E_R+E_L+h_{E_R, E_L} \right)}{8 E_RE_L}\right)^{\f{1}{d+1}},
\end{align}
\begin{align}
 \phi_s = \f{\pi}{2}+i \log\sqrt{\f{d-1}{2(d+1) E_L} \left(E_R-E_L+h_{E_R, E_L}  \right)}\,,
 \end{align}
 \begin{align}
\textrm{with }\hspace{2mm} h_{E_R, E_L} = \sqrt{E_R^2+\f{2(d^2+6d+1)}{(d-1)^2} E_RE_L+ E_L^2}\,.
\end{align}
Notice that the saddle-point value $\phi_s$ is independent of $E_{\textrm{vac}}$, which indicates a theory-independent universality as $\tau$ is varied and Im$[\tau]/$Re$[\tau]$ is kept fixed.

Plugging these saddles into the integrand gives us the following expression for the entropy
\begin{align}
\log \rho(E,J) = \pi\sqrt{d+1} \left(\f{2}{d^d}\right)^{\f{1}{d+1}} \left(\sqrt{(d+1)^2 E^2-4 d J^2}-(d-1)E\right)^{\frac{d-1}{2 (d+1)}}\nonumber\\
\times \left(\sqrt{(d+1)^2 E^2- 4d J^2}+ (d+1)E\right)^{1/2} (-E_{\textrm{vac}})^{\f{1}{d+1}}\,.\label{rotatingentropy}
\end{align}
For a gapped spectrum with vacuum  satisfying $E_R=E_L=E_{\textrm{vac}}/2$, we have $\widetilde{Z}(-r_s^{-1},-\phi_s)^{r^{1-d}}\approx 1$, as required for consistency of the saddle-point approximation.

As expected, the degeneracy of states does not factorize into left-moving and right-moving pieces unless $d=1$. Even the first correction in $d=1+\varepsilon$ does not factorize:
\be
\log \rho(E_R, E_L)=\sqrt{-E_{\textrm{vac}}/2}\left(4\pi\left(\sqrt{E_R}+\sqrt{E_L}\right)+\varepsilon\,\f{\pi}{2}\left(\sqrt{E_R}+\sqrt{E_L}\right) \log\left(\f{E_R\, E_L}{-E_{\textrm{vac}}}\right)+\dots\,\right).
\ee

\subsection{Angular momentum on the torus: new-school derivation}\label{take2}
The previous derivation was presented only to show that techniques familiar from two dimensions can be utilized. In the thermodynamic limit, the degeneracy of states with angular momentum can more easily be obtained from the case without angular momentum by a Lorentz transformation. 

Since the entropy is invariant under boosting, our answer should not change. However, we are not taking into account the length contraction factor when comparing the boosted case to the unboosted case, since the finite-size spatial cycle is the same in both cases. This means the two entropy results should differ by a length contraction factor $\gamma=1/\sqrt{1-a^2}$. This is of course true for a boost in an arbitrary direction, where in that case $a^2 = \sum a_i^2$. Thus in the regime $S\propto V_d$ and the thermodynamic limit where $S=\log \rho$ we have
\be
\log \rho(E,J)= \f{\log \rho(E_{\textrm{nr}}, J=0)}{\sqrt{1-a^2}}\,.\label{quickentropy}
\ee 
 This is a formula for the degeneracy of states in terms of the non-rotating energy $E_{\textrm{nr}} = E(a=0)$ and velocity $a$. To obtain the degeneracy in terms of the rotating energy and angular momentum, we can solve for $a$ and $E_{\textrm{nr}}$ in terms of $E$ and $J$ using the Lorentz transformation law of the stress-energy tensor $T^{\mu\nu} = \Lambda^{\mu}_{\alpha} \Lambda^{\nu}_{\beta} T^{\alpha\beta}_{\textrm{nr}}$:
\be
E=E_{\textrm{nr}}\left(\f{1+a^2/d}{1-a^2}\right), \qquad J=E_{\textrm{nr}}\,a\left(\f{1+1/d}{1-a^2}\right)\,.\label{transfers}
\ee
We have kept the size of the spatial circle the same in both the rotating and non-rotating cases when integrating the densities $T^{00}$ and $T^{01}$ to get the energy and angular momentum. Plugging the solutions for $a$ and $E_{\textrm{nr}}$ into the right-hand-side of \eqref{quickentropy} results in the expression for $\rho(E,J)$  given in \eqref{rotatingentropy}. 
This provides a new derivation of the Cardy formula at finite angular momentum for $(1+1)$-dimensional CFTs.

\subsubsection{Hartman-Keller-Stoica conjecture}
In the case of $(1+1)$-dimensional CFTs, Hartman, Keller, and Stoica (HKS) showed that the Cardy degeneracy is universal for all states with $E_{\textrm{nr}}>c/12$ as long as the degeneracy of states with $E_{\textrm{nr}}<0$ is bounded as $\rho(E_{\textrm{nr}})\lesssim \exp(2\pi(E_{\textrm{nr}}+c/12)$.\footnote{Thanks to Tom Hartman for discussions about HKS.} For the rotating ensemble, they conjectured that the Cardy degeneracy is universal for $E_L E_R>(c/24)^2$ as long as (in addition to the previous bound) the degeneracy of light states (i.e. states with either $E_L$ or $E_R$ negative) is bounded as $\rho(E_L,E_R) \lesssim \exp[4\pi \sqrt{(E_L+c/24)(E_R+c/24)}]$. Notice that we can produce any state in the range $E_L E_R > (c/24)^2$ simply by boosting a static state $E_L=E_R>c/24$. By consistency with local Lorentz invariance, we therefore conclude that states in this range are universal without any assumptions beyond the ones needed in the static ensemble. Let us be a bit more explicit.

We begin by using the result of HKS that at zero angular potential, the Cardy degeneracy $\rho(E_{\textrm{nr}})=\exp\left(2\pi \sqrt{cE_{\textrm{nr}}/3}\right)$ for central charge $c$ is universal in the range $E_{\textrm{nr}}>c/12$, as long as the degeneracy of states with $E_{\textrm{nr}}<0$ satisfies $\rho(E_{\textrm{nr}})\lesssim \exp\left(2\pi(E_{\textrm{nr}}+c/12)\right)$. The ingredients we will use are that the entropy and stress tensor arising from the analysis of \cite{Hartman:2014oaa} transform covariantly under Lorentz transformations, and that the states with universal degeneracies can be smoothed into a density of states which satisfies $\log \rho(E_L, E_R) = S(\beta_L, \beta_R)$. In particular, the stress-tensor transforms as $T_{\mu\nu}' = T_{\alpha \beta} \Lambda^{\alpha}_{\mu} \Lambda^{\beta}_{\nu}$ and the entropy  in the universal range is invariant under boosts. These are our assumptions, which can be violated for otherwise healthy CFTs.

Since Lorentz invariance is broken globally, there is a preferred observer. Physics which probes the global topology will not be relativistically invariant. Alice can wind around the circle in a \emph{fixed} inertial frame but still return younger than her twin Bob, who is idly waiting in the preferred rest frame. For our purposes, the effect of this global breaking of Lorentz invariance will be that the size of the universe changes under boosts, which is the source of the factor of $\gamma$ in equation \eqref{quickentropy}.

Using \eqref{quickentropy} and \eqref{transfers}, we can translate results from the static ensemble to the rotating ensemble. For a spectrum with the constraint
\be
\log\rho(E<0, J=0) \lesssim 2\pi (E+c/12)
\ee
we have
\be\label{boostanswer}
\log \rho(E_L,E_R) = 2\pi \sqrt{cE_L/6}+2\pi\sqrt{cE_R/6},\qquad E_{\textrm{nr}}=\sqrt{E_LE_R}>c/24\,.
\ee
Notice that in the entropy formula \eqref{boostanswer} we have $E_L = L_0 - c/24$, $E_R = \bar{L}_0 - c/24$. In particular, the additive factors of $-c/24$ are correctly obtained, even though one might expect that this boosting procedure should only work for $L_0$, $\bar{L}_0 \gg c$, where the size of the compact circle can be neglected. 

Boosting states from the static ensemble is not a panacea; the ordering (i.e. whether $E<J$ or $E>J$) is maintained under boosts. In particular, one cannot obtain states with $E_R<0$ and $E_L>0$ (or vice versa) from boosting states with $E_L=E_R$. Also, it is crucial that we are boosting in a region of the spectrum for which $\rho(E_L,E_R) \sim e^{S(\beta_L,\beta_R)}$, i.e. we can average to produce a smooth approximation to the density of states. For example, we cannot boost the vacuum state and use the above technology. 

Notice that our assumptions \emph{imply} a sparseness condition on the light states with $E_L E_R<(c/24)^2$ with $E_L$ and $E_R$ the same sign. (States with $E_L$ and $E_R$ of opposite sign have $|J|>E$ and it is not clear how our formalism applies to these states. Naively, they require boosts with $a>1$ to reach, and our formula \eqref{quickentropy} goes haywire in this faster-than-light regime.) In particular, if there are many states in this range, allowing a smoothed approximation that we could then boost, we would run into a contradiction by boosting to a state with $E<0$ and $J=0$, whose degeneracy \emph{is} constrained by the assumptions above. We can get an estimate for this sparsness condition by assuming we have a density of states we can boost and finding the bound in the boosted frame:
\begin{align}
\rho(E<c/12, J=0) &\lesssim 2\pi(E+c/12) \\
\longrightarrow \quad \rho(E_L,E_R;E_LE_R<(c/24)^2)&\lesssim \frac{\pi  \left(\sqrt{E_L}+\sqrt{E_R}\right) (c+12 (E_L+E_R))}{12 (E_LE_R)^{1/4}}\,.
\end{align}

When trying to boost a state with angular momentum to another state with higher or lower angular momentum, one cannot do the procedure exactly as above. Since the zero angular momentum state is in the preferred rest frame of the circle, one should first boost to this state (this time picking up a factor of $1/\gamma$ instead of $\gamma$ in the entropy; this asymmetry is due to breaking global Lorentz invariance) and then boost to the state with higher or lower angular momentum. 

\subsection{Boosted black branes}\label{boostbrane}
The boosted black brane can be written as \cite{Hubeny:2011hd}
\begin{align}
ds^2=\ell^2\left(-f(r)u_\mu u_\nu dx^\mu dx^\nu+r^2\left(u_\mu u_\nu+\eta_{\mu\nu}\right)dx^\mu dx^\nu+\f{dr^2}{f(r)}\right),\\
u_\mu = \left(\f{-1}{\sqrt{1-a^2}},\,\f{a}{\sqrt{1-a^2}},\,\vec{0}\right), \qquad f(r) = r^2\left(1-\left(\f{r_h}{r}\right)^{d+1}\right)\,.
\end{align}
We have chosen the velocity to be in a single direction, although this can be generalized. 

We can now compactify one of the directions onto a finite-size circle. The mass and angular momentum can be obtained from the renormalized Brown-York stress tensor, and the entropy is given as usual by a quarter of the horizon area in Planck units:
\be
E = \frac{\left(d+a^2\right) V_d\,\ell^d\, r_h^{d+1}}{16\pi G \left(1-a^2\right) }, \qquad J = \frac{(d+1) V_d \, \ell^d \,r_h^{d+1}\,a}{16\pi G (1-a^2)}, \qquad S_{\textrm{BH}} = \f{A}{4G} = \f{ V_d \,\ell^d\,r_h^d}{4G\sqrt{1-a^2}\,}\,.\label{boostentropy}
\ee
$V_d$ represents the volume of the spatial torus of the field theory. These can also be obtained as in the previous subsection by the Lorentz transformation law on the stress-energy tensor of the unboosted black brane $T_{\textrm{boosted}}^{\mu\nu} = \Lambda^{\mu}_{\alpha} \Lambda^{\nu}_{\beta} T^{\alpha\beta}_{\textrm{static}}$.

Plugging in this energy, angular momentum, and the vacuum energy from the AdS soliton \eqref{generald} into \eqref{rotatingentropy} gives precisely the Bekenstein-Hawking entropy calculated above:
\be
S_{\textrm{BH}} = \log \rho(E,J) \,.
\ee

\subsection{Angular momentum on a sphere}\label{sphereft}
When a fluid dynamical regime is valid, the partition function of a CFT$_{d+1}$ on the sphere at finite temperature and angular potential can be obtained from the non-rotating case as \cite{Bhattacharyya:2007vs}
\be
\log Z_{\textrm{rotating}} = \f{\log Z_{\textrm{nr}}}{\prod_i (1-\Omega_i^2)}.\label{fluid}
\ee
The fluid regime is  applicable because we are working in the thermodynamic limit in the Cardy regime, i.e. at large energy density/temperature, and at strong coupling.

In this case we have picked the most general rotation possible, i.e. we have $\left \lfloor{ (d+1)/2}\right \rfloor$ independent rotation parameters. The universal rotation-dependent factor carries through to the thermodynamic entropy and the microcanonical degeneracy of states. To see it in the thermodynamic entropy, notice that 
\begin{align}
\left(-\beta \p_\beta - \h_j \p_{\theta_j}\right) &\;\f{1}{\prod_j 1-\Omega_j^2} = 0,\qquad \textrm{since } \h_j = i \beta \Omega_j\\
\implies S=\left(1-\beta \p_\beta -\h_j \p_{\theta_j}\right)&\log Z = \f{(1-\beta \p_\beta)\log Z_{\textrm{nr}}}{\prod_i(1-\Omega_i^2)}=\f{S_{\textrm{nr}}}{\prod_i(1-\Omega_i^2)}
\end{align}
To see the prefactor carry through into the microcanonical degeneracy of states, we consider the inverse Laplace transform which takes us from the partition function to the microcanonical degeneracy of states:
\be
\rho(E,J_1, J_2,\dots, J_n) = \int d\beta \left(\prod_i d\Omega_i\right) Z(\beta,\Omega_1, \Omega_2, \dots, \Omega_n)\,e^{\beta (E-\Omega_i J_i)}\,.
\ee
Using the expression for the partition function \eqref{fluid}, we can write this as 
\be
\rho\left(E,J_1, J_2, \dots, J_n\right) = \int d\beta  \left(\prod_i d\Omega_i\right)  Z(\beta)^{\f{1}{\prod_i \left(1-\Omega_i^2\right)}}\,e^{\beta(E-\Omega_i J_i)}\label{step2}
\ee
The most elegant way to proceed is to replace the partition function on the sphere with the partition function on the torus $\mathbb{T}^2 \times \mathbb{R}^{d-1}$, which enjoys the modular properties previously discussed:
\be
Z(\beta) =Z\left(\f{L^{d+1}}{\beta^d}\right)^{(L/\beta)^{d-1}}\,.
\ee
Plugging the modularly transformed partition function into \eqref{step2} and evaluating the integral by saddle-point approximation for large $E$ yields saddles for $\beta=T^{-1}$ and $\Omega_i$ implicitly defined by 
\begin{align}
E&=\f{V_d T^{d+1} \varepsilon_{\textrm{vac}}}{\prod_a\left(1-\Omega_a^2\right)} \left(\sum_a \f{2\Omega_a^2}{1-\Omega_a^2}+d\right),\label{energy}\\
J_i &= \f{V_d T^{d+1} \varepsilon_{\textrm{vac}}}{\prod_a \left(1-\Omega_a^2\right)} \left(\f{2\Omega_i}{1-\Omega_i^2}\right).\label{angular}
\end{align}
These saddle-point values are in agreement with the thermodynamic energy and angular momenta given in \cite{Bhattacharyya:2007vs}. We now see that the constant $h$ in \cite{Bhattacharyya:2007vs} equals $\varepsilon_{\textrm{vac}} = -E_{\textrm{vac}} L^{d+1}/V_d$, where $V_d=\textrm{Vol}( S^d)=2\pi^{(d+1)/2}/\Gamma[(d+1)/2]$. 

The microcanonical degeneracy of states at energy $E$ and angular momenta $J_i$ can be given implicitly in any number of dimensions as 
\be
\log \rho(E,J_1,J_2,\dots,J_n) = \f{(d+1)T^d V_d \varepsilon_{\textrm{vac}}}{\prod_i (1-\Omega_i^2)}\,.\label{canonical}
\ee
There can be at most $N=\left\lfloor{(d+1)/2}\right\rfloor$ independent rotation parameters. The expression in terms of the energy $E$ and angular momenta $J_i$ can be written explicitly by solving \eqref{energy}-\eqref{angular} for $T$ and $\Omega_i$ in terms of $E$ and $J_i$ and plugging into \eqref{canonical}. By the state-operator correspondence of CFTs on the sphere, this degeneracy can be understood as the degeneracy of local operators with scaling dimension $E$ and spins $J_i$.

The explicit expression for the degeneracy in terms of the energy and angular momenta is generally messy, although there are some cases where it simplifies. With a single rotation parameter, in any number of dimensions, we have 
\be
\log \rho(E,J)=\f{(d+1)(\varepsilon_{\textrm{vac}}V_d)^{\f{1}{d+1}}}{2^{\frac{1}{d+1}}  d^{\frac{d}{d+1}}(d-2)^{\frac{d-2}{d+1}}}  \left(\frac{-J^2 \left((d-1) E-\sqrt{(d-2) d J^2+E^2}\right)^d}{(d-2) J^2+E^2-E \sqrt{(d-2) d J^2+E^2}}\right)^{\frac{1}{d+1}}\,.
\ee
 The expression is particularly simple in $d=2$:
\be
\log\rho(E,J) =\f{3}{2^{2/3}} \left(\varepsilon_{\textrm{vac}} V_2\right)^{1/3} \left(E^2-J^2\right)^{1/3} \,.
\ee
We also state the result in even dimensions where all rotation parameters are turned on and equal. We have $d/2$ directions of rotation, and we find
\begin{align}
\log \rho(E,J) =& \f{d+1}{d^{d/(d+1)}}(\varepsilon_{\textrm{vac}}V_d)^{\f{1}{d+1}}\left(E^2-\f{d^2}{4}\,J^2\right)^{\f{d}{2(d+1)}}\\
\implies& \log \log \rho(E,J)\sim\log E_++\log E_-\,.
\end{align}
We have defined $E_{\pm} = E \pm dJ/2$ to exhibit a fascinating factorization into ``left-moving" and ``right-moving" contributions. This factorization is in the logarithm of the logarithm of the degeneracy, as opposed to the case of $d=1$ where the logarithm of the degeneracy factorizes.  The regime of applicability of the formula is $E_\pm \rightarrow \infty$. It is not clear what physical property this factorization is indicating.

Notice that we cannot freely pick both the energy and the angular momentum in these expressions. For example, $E=dJ/2$ would indicate a vanishing entropy, but this is not an acceptable limit of our formula. This is most simply explained in the thermodynamic limit in which we are operating. In that case, \eqref{energy}-\eqref{angular} make it clear that as $E/(dJ/2)\rightarrow 1$, we have $\Omega\rightarrow 1$ and so $E-dJ/2\rightarrow \infty$. The easiest way to understand the regime of applicability of the formula is in terms of the implicit formula \eqref{canonical}, for which the temperature is large and $0<\Omega<1$.


\subsubsection{Universal quantities}
 The ratio of the log of the number of local operators at large and fixed $E$ and spins $J_i$ with the log of the number of local operators at $E$ and spins $\tilde{J}_i$ is theory-independent (within the class of theories we are considering):
\be
\f{\log \rho(E,J_i)}{\log \rho(E,\tilde{J}_i)} = \f{\prod_i (1-\widetilde{\Omega}_i^2)}{\prod_i (1-\Omega_i^2)}\,.
\ee
In $1+1$ dimensions, we have $\log \rho(E,J) = \left(\varepsilon_{\textrm{vac}} V_1\right)^{1/2}(\sqrt{E+J}+\sqrt{E-J})$, which we rewrite in the form of \eqref{canonical} and obtain 
\be
\f{\log \rho(E,J)}{\log \rho(E,\tilde{J})} = \f{1-\widetilde{\Omega}^2}{1-\Omega^2}\,.
\ee
In this case we have $J=2\Omega E/(1+\Omega^2) $. We see again that as $J\rightarrow E$, the ratio diverges. But notice that we are at $T>0$, and this cannot be interpreted as an extremal limit.

\subsection{Kerr-AdS$_{d+2}$ black hole}\label{spheregravity}
The general Kerr-AdS$_{d+2}$ metric, with up to $n=\left \lfloor{(d+1)/2} \right \rfloor$ independent rotation parameters, was first presented in \cite{Gibbons:2004uw}. The energy and angular momenta can be computed relative to a frame that is non-rotating at infinity \cite{Gibbons:2004ai}, and for large horizon-radius one finds precisely our saddle-point values \eqref{energy} and \eqref{angular}. The entropy of the hole with large horizon-radius is given as 
\be
S_{\textrm{BH, Kerr}} = \f{S_{\textrm{BH, Schwarzschild}}}{\prod_i (1-\Omega_i^2)}\,, \qquad i=1,2,\dots,n\,.
\ee
This precisely mirrors our implicit formula for the microcanonical degeneracy of states
\be
\log \rho(E, J_1, J_2,\dots,J_n) = \f{\log \rho(E(\Omega_i = 0))}{\prod_i (1-\Omega_i^2)}\,.
\ee
This establishes that the Bekenstein-Hawking entropy of a large Kerr-AdS black hole with arbitrary angular momenta is given as the microcanonical degeneracy of states (or local operators, by the state-operator correspondence) of the dual field theory. The numerical coefficient of the entropy is controlled by the vacuum energy of the theory on $S^1\times\mathbb{R}^{d-1}$ with supersymmetry-breaking boundary conditions along the $S^1$.

\section{Generalizations}
\subsection{$U(1)$ charge}
It is well-known that for two-dimensional CFTs, extending the partition function to include states charged under a global $U(1)$ symmetry modifies the modular properties. The partition function for this ensemble, defined as
\be
Z(\tau,\bar{\tau},z,\bar{z}) =\textrm{Tr}\,\left(e^{2\pi i \tau (L_0-\hat{c}/24)}e^{-2\pi i \bar{\tau}(\bar{L}_0-\hat{\bar{c}}/24)}e^{2\pi i z Q}e^{-2\pi i \bar{z}\bar{Q}}\right),
\ee
 becomes a Jacobi form of weight $0$ and index $k$:
 \be
 Z(\tau,\bar{\tau},z,\bar{z})  = e^{-2\pi i k \f{cz^2}{c\tau+e}}e^{2\pi i k \f{c\bar{z}^2}{c\bar{\tau}+e}}Z\left(\f{a\tau+b}{c\tau+e},\f{a\bar{\tau}+b}{c\bar{\tau}+e}, \f{z}{c\tau+e}, \f{\bar{z}}{c\bar{\tau}+e}\right).
 \ee
The central charges are denoted $\hat{c}, \hat{\bar{c}}$. The level $k$ is meaningful in cases where the $U(1)$ symmetry is embedded into a larger symmetry group, like an $SU(2)$ algebra or the $\mathcal{N}=2$ superconformal algebra.

One way to derive this transformation law is to refer to the free boson and perform the Legendre transform which takes you from the path integral to the trace over the Hilbert space. The path-integral representation is modular invariant, but picks up a term quadratic in the potential $z$ when performing the Legendre transform. This term transforms under a modular transformation, and to cancel it requires the prefactor written above. In two dimensions, $U(1)$ symmetries can always be represented in terms of free bosons, so this argument is general (see \cite{Kraus:2006wn} for details). One could also derive this transformation from the path integral by carefully dealing with the singular operator product expansion of the current $J$ with itself.

Although we do not have a general derivation, we conjecture that in higher dimensions the partition function on $\mathbb{T}^2 \times \mathbb{R}^{d-1}$ picks up the same sort of anomalous prefactor quadratic in the potential, in addition to a spatial volume factor which rescales by $(L/\beta)^{d-1}$. This indeed happens for a free boson in higher dimensions, though this theory is not a CFT. Schematically we have
\be\label{proposal}
Z \sim e^{-(z^2/\beta)(L/\beta)^{d-1}(V_{d-1}/L^{d-2})} Z'\,,
\ee
where $Z'$ is the modularly transformed partition function and $z \sim \beta \Phi$ for chemical potential $\Phi$. The factor of $L^{d-2}$ has been put in the exponent to make it dimensionless and has no obvious source, a point which we will return to later. 

With this proposed modular property, we can perform the inverse Laplace transform explicitly for $d=2$ and $d=3$. In an expansion for small charge $Q$, we find that the entropy picks up the following contribution:
\be
\log \rho(E, Q) \sim E^{\f{d}{d+1}}V_d^{\f{1}{d+1}}-E^{\f{d}{d+1}}V_d^{\f{1}{d+1}}\sum_{n=1}^{\infty}c_n\left(\f{Q}{E^{\f{d}{d+1}}V_d^{\f{1}{d+1}}}\right)^{2n};\qquad d=2,3\,.
\ee
In the above expression we have assumed the ground state has vanishing charge (so we are fixing the chemical potential and summing over the charges). This functional dependence on the charge agrees with Reissner-Nordstrom-AdS black branes in the bulk, although the coefficients $c_n$ do not agree. 

By the technology in section \ref{take2}, we can obtain the entropy at finite angular momentum on the torus and match to boosted Reissner-Nordstrom-AdS black branes. We can also consider angular momentum and charge on the sphere by utilizing the following result for conformal fluids on a sphere \cite{Bhattacharyya:2007vs}:
\be
\log Z(\beta,\Phi,\Omega_i) = \f{\log Z(\beta,\Phi)}{\prod_i (1-\Omega_i^2)}.\label{fluidgen}
\ee
The universal angular-velocity factors carry through to the degeneracy of states as we saw in \ref{sphereft}. This agrees with the entropy of the Kerr-Newman black hole at high temperature,  which equals the entropy of the Reissner-Nordstrom black hole with the same universal angular-velocity factors.

It would be interesting to derive from first principles the modular properties for a higher-dimensional CFT with a $U(1)$ charge to check the statements above. One potentially confusing point about the proposal above arises from dimensional analysis. For the free boson in higher dimensions, the current $\partial_\mu \phi$ has dimension $(d+1)/2$, which means the coupled source $A^\mu$ has dimension $(d+1)/2$. The factor $\int d^{d+1}x \,A_0^2$ is therefore dimensionless and is a natural object to appear in the exponent. This is in fact what appears when performing the transform. The problem arises when considering a CFT. In that case, a spin-$1$ current $J_\mu$ has dimension $d$, meaning the coupled source $A^\mu$ has dimension $1$. This means something has to make up the dimensions in the type of term we have postulated, $e^{\int A_0^2}$. (We have simply put in additional factors of $L$ to make up the dimensions.) Notice that for $d=1$ we have $(d+1)/2=d$, so this issue does not arise. This is because the free boson in this case is a CFT. To see the Legendre transform in action, we can take the complex, conformally coupled scalar in higher dimensions as our representative example. It has $U(1)$ current $J_\mu = i (\phi \partial_\mu \phi^{\dagger}-\phi^{\dagger} \partial_\mu \phi)$. If we source $J_0$ in the path integral with $A_0 J^0$, then the canonical momenta are $\pi_{\phi} = \partial_t \phi^{\dagger}- i A^0\phi^{\dagger}$ and its complex conjugate. This leads to a quadratic term $e^{\int A_0^2|\phi|^2}$ in the relation between the path integral and trace formula. So, the dimensions are made up by $|\phi|^2$. Although this looks like a mass term which breaks scale invariance, the ``mass" $A_0$ is spurionic and transforms under a scale transformation. Together with the piece $A_0 J^0$, these terms have the effect of shifting the Matsubara frequencies by a constant imaginary term. In particular, the free energy remains extensive, which justifies our additional factor of $(L/\beta)^{d-1}$. Notice that the piece $\int A_0^2 |\phi|^2$ cannot be factored out as an anomalous prefactor as in \eqref{proposal}. Instead, it enters into the Hamiltonian. Nevertheless, we know in two dimensions that the end result is the behavior \eqref{proposal}. This emphasizes the necessity of proving the correct anomalous dependence in higher dimensions in a more general way. One may suspect that the procedure of regularizing contact terms in the conserved current operator product expansion, as can be done in two dimensions, will lead to a general answer. However, in the case of e.g. a Dirac fermion in $2+1$ dimensions, there does not seem to exist such an anomaly \cite{Hsieh:2015xaa}.\footnote{Thanks to Shinsei Ryu for explaining this case to me and to Gabor Sarosi for discussions about a general derivation.}

\subsection{Hyperscaling violation and D$p$-branes}
In section \ref{euclidean}, we showed that the localization of the field theory partition function on the ground state can be understood from the bulk as a localization of the Euclidean gravity partition function on the black brane configuration. This is general and remains true in any context where (a) the field theory partition function projects to the ground state and (b) the ground state configuration in the bulk is given by a soliton obtained by double Wick rotation from the black brane. As shown in \cite{Shaghoulian:2015kta}, this projection remains true once we add a hyperscaling-violation exponent $\h$ into the mix as long as the specific heat of the black brane remains positive ($\h<d$). Thus, since Euclidean gravity methods give the correct entropy, formula \eqref{microeff} will also reproduce the correct entropy. This has been worked out explicitly for $d=1$ in \cite{Shaghoulian:2015dwa}. We omit the details of the higher-dimensional calculation, but it straightforwardly follows the approach of the previous sections and shows $S_{\textrm{BH}} = \log \rho(E)$. Boosted branes can also be constructed, with their entropy given most simply as in section \ref{take2}. 

Black D$p$-branes, when dimensionally reduced over the sphere, give rise to hyperscaling-violating black brane solutions with particular values of $\h$ \cite{Dong:2012se}. These are special cases of our derivation for general $\h$. 

\subsubsection{D$5$-branes}
One of the more remarkable examples of holographic duality is derived from considering a stack of NS$5$-branes (or the S-dual D$5$ branes). The worldvolume theory is exotic and dubbed ``little string theory" for its similarity to string theory. More precisely, in the decoupling limit necessary to derive the duality, $\alpha'$ is kept finite on the brane while $G_N\rightarrow 0$. Thus there is no gravity on the background but there are still strings. When compactified on tori the theory enjoys T-duality as a \emph{symmetry}, and the density of states can be shown to be of Hagedorn type. 

The gravitational background is given as 
\be
ds^2 =\ell^2r^2\left(-(1-r_h^2/r^2)dt^2+\f{dr^2}{r^2(1-r_h^2/r^2)}+dx_i^2\right)\qquad i=1,...,5\,,
\ee
which can be obtained by sending $\h\rightarrow -\infty$ \cite{Shaghoulian:2013qia}. This makes the effective dimensionality diverge, an indication of the stringyness of the dual. The inverse temperature of this black brane is $\beta = 2\pi$, independent of the location of the horizon. This allows us to integrate $dE = TdS$ to get $S = 2\pi E$. The entropy formula \eqref{microeff} in the limit $\h\rightarrow -\infty$ becomes 
\be
S = 2\pi E\,,
\ee
where we fixed the length of the spatial cycle to $2\pi$ and $E$ is the energy of the black brane. We have also made the strong assumption that $\log E_{vac}\sim d_{\textrm{eff}}^n$ for $n<1$ so that $\lim_{d_{\textrm{eff}}\to \infty} (-E_{\textrm{vac}})^{1/(d_{\textrm{eff}}+1)} = 1$. This is not a sensible limit since we do not have a family of theories indexed by $d_{\textrm{eff}}$, and even in (irrelevant) cases for which we do have a family of theories--like a massless scalar field--it is violated. Nevertheless, the final formula matches the bulk thermodynamics, and it can be obtained by more honest means. Proceeding as in \cite{Shaghoulian:2015dwa}, we can do a fresh analysis of the partition function. We assume that the symmetry of this spacetime $r\rightarrow \lambda r$, $\ell\rightarrow \lambda \ell$ is inherited by the gravitational sector of the dual theory as $E\rightarrow \lambda E$. Then we see
\be
Z(\beta) = \int dE \,e^{-\beta E} \rho(E) = \int (\lambda dE) e^{-\lambda \beta E} \rho(\lambda E)\implies \rho(E) = e^{\beta E}(\delta(E)+1/E)\,.
\ee

\section{Logarithmic corrections}

\subsection{Gravity}
In \cite{Sen:2012dw}, it was argued that there are three possible sources of logarithmic corrections to the entropies of black holes. Two of these sources are zero  modes and nonzero modes, which enter into the partition function directly:
\be
Z = Z_{\textrm{nz}} + Z_{\textrm{zm}}\,.
\ee
These corrections from zero modes and nonzero modes and can be calculated as a one-loop determinant in the Euclidean gravity path integral which defines the partition function. 

The final source of logarithmic corrections arises from transforming the partition function into the microcanonical density of states. For the general hyperscaling-violating black brane, we find 
\be
S_{\textrm{BH}} = \f{A}{4G}-\f{2+d_{\textrm{eff}}}{2d_{\textrm{eff}}} \log \f{A}{4G}\,.\label{bulklog}
\ee

\subsection{Field theory}
Just as in the bulk, the field theory has logarithmic corrections which come from the inverse Laplace transform from the partition function to the microcanonical density of states. The transform can be calculated by a saddle-point approximation as before, but this time we keep the first correction to the leading saddle. We have
\be
\rho(E) = \int d\beta \,Z(\beta) \,e^{\beta E}\implies\log \rho(E) \approx S_0-\f{2+d_{\textrm{eff}}}{2\,d_{\textrm{eff}}} \log S_0\,.\label{logcorr}
\ee
The logarithmic correction is determined entirely by the specific heat and reproduces \eqref{bulklog}. To illustrate this derivation in a bit more detail, we restrict to a scale-invariant theory and use the modular invariance on $\mathbb{T}^2 \times \mathbb{R}^{d-1}$:
\be
\rho(E) = \int d\beta \,Z(\beta) \,e^{\beta E}= \int d\beta \, \left(\widetilde{Z}\left(L^2/\beta\right)\right)^{(L/\beta)^{d-1}}\left(e^{L^2 E_{\textrm{vac}}/\beta}\right)^{(L/\beta)^{d-1}} e^{\beta E}\,,
\ee
where we defined $\widetilde{Z}(\beta) := e^{-\beta E_{\textrm{vac}}}Z(\beta)$ and shifted the definition of the energies to mimic the usual two-dimensional derivation. This integral is evaluated by saddle point on the part of the integrand independent of $\widetilde{Z}$, and it can be checked that $\widetilde{Z}(L^2/\beta^*)= 1$ on the saddle $\beta^*$. We find
\be
 \log \rho(E) \approx S_0 -\f{2+d}{2\,d} \log S_0\,.
\ee
This derivation can be generalized to incorporate hyperscaling-violation, and reproduces \eqref{logcorr}. 

Notice that since the field theory and bulk partition functions localize to the same contribution at high temperature as argued in section \ref{euclidean}, they are forced to give the same logarithmic corrections. Projecting the partition function at high temperature before inverse Laplace transforming is sensible if one assumes the dominant saddle is at high-temperature. Of course, as we showed above, this assumption is not necessary and the modular properties can be used to provide a more rigorous treatment. 

What about logarithmic corrections which are analogous to the zero modes and nonzero modes in the bulk? In our formalism, such corrections enter into the vacuum energy. This is clear from the bulk, since the vacuum energy is calculated by a Euclidean gravity partition function. Since the Euclidean geometries of the black brane and the vacuum state match (up to rescalings), any logarithmic corrections will trivially match between the two. 

\section{Discussion}\label{conclusions}
By using the Casimir energy implied by the AdS soliton and its hyperscaling-violating cousins, we have been able to use formula \eqref{microeff} to give a microscopic count of the states of various black holes and black branes. This is the natural generalization of Strominger's calculation of the BTZ black hole entropy to higher dimensions and to violations of hyperscaling.

The entropy formulas (and other thermodynamic formulas) derived above exhibit a universal appearance of the vacuum charge $\varepsilon_{\textrm{vac}}$. In the case of $\mathcal{N}=4$ SYM, this number differs by a factor of $3/4$ between the weak and strong coupling results. This of course feeds into the famous $3/4$ difference between the thermal entropy at weak and strong coupling. The appearance of a factor of $3/4$ was also observed in various thermodynamic formulas in extended ensembles by \cite{Hawking:1999dp, Berman:1999mh, Landsteiner:1999xv}, but was left unexplained. We see that the universal appearance of the vacuum charge in the various formulas above explains the universal appearance of this factor.

The torus partition function in higher dimensions does not have a direct relationship with the spectrum of local operators. In theories like pure Chern-Simons theory, it is instead related to the spectrum of line operators.\footnote{Thanks to David Simmons-Duffin for discussions about this case.} Modular invariance in this case is not interesting, since the theory is topological and therefore independent of cycle lengths. However, it would be interesting to investigate conformal Chern-Simons-matter theories, which play a key role in condensed matter physics and holography for four-dimensional spacetimes. 

We have only used our various modular properties for asymptotically small $\beta$, and it would be interesting to explore the constraints it provides at intermediate $\beta$. For example, one could utilize the modular bootstrap invented in \cite{Cardy:1991kr} and applied to quantum gravity in \cite{Hellerman:2009bu}. This would give information about states on the background $\mathbb{T}^2\times \mathbb{R}^{d-1}$. The Chern-Simons-matter theories mentioned above are very sensitive to topology. In particular, new states are introduced on nontrivial backgrounds, including $\mathbb{T}^2\times \mathbb{R}$. This higher-dimensional modular bootstrap may therefore elucidate the role of topology in four-dimensional quantum gravity and cosmology \cite{Banerjee:2012gh, Banerjee:2013mca}.

Another fascinating application of this formalism would be to non-unitary theories. The derivation of the Cardy formula uses unitarity by considering a spectrum of real energies bounded below. For CFTs in two dimensions, considering non-unitary theories may connect to the entropy of the cosmological horizon in dS$_3$ \cite{Bousso:2001mw, Anninos:2009yc, deBuyl:2013ega}. Using bulk-inspired assumptions to constrain the dual field theory, one may hope to derive a Cardy formula for these special theories. This approach has many interesting subtleties \cite{medio}. However, for dS$_4$, there exist concrete proposals for non-unitary CFT duals \cite{Anninos:2011ui, Anninos:2013rza, Chang:2013afa, Anninos:2014hia}. The non-unitary nature of these theories is much milder, since the weights of local operators are real and non-negative, and our formalism should apply. One could then investigate whether the degeneracies of states in these theories have any relation to the entropy of the cosmological horizon in the bulk.

\section*{Acknowledgments}
I would like to acknowledge useful discussions with Dionysios Anninos, Tarek Anous, Steven Carlip, William Donnelly, Tom Hartman, Gary Horowitz, Per Kraus, Aitor Lewkowycz, David Mateos, Shinsei Ryu, Gabor Sarosi, David Simmons-Duffin, Tomonori Ugajin, and Huajia Wang. I would like to particularly thank Dionysios Anninos for Cardy conversations and comments. This work is supported by NSF Grant PHY13-16748.

\bibliography{cardyhigherbiblio}
\bibliographystyle{apsrev4-1long}

\end{document}